\input{aipcheck}

\documentclass[
    ,final 
  ]
  {aipproc}

\layoutstyle{6x9}

\begin{document}

\title{On the nonextensive character of some magnetic systems}



\author{D. O. Soares-Pinto}{
  address={Centro Brasileiro de Pesquisas
F\'{\i}sicas, Rua Dr. Xavier Sigaud 150, 22290-180, Rio de
Janeiro, Brazil}
  ,altaddress={CICECO, Universidade de Aveiro, 3810-193,
Aveiro, Portugal} 
}

\author{M. S. Reis}{
  address={CICECO, Universidade de Aveiro, 3810-193,
Aveiro, Portugal} }

\author{R. S. Sarthour}{
  address={Centro Brasileiro de Pesquisas
F\'{\i}sicas, Rua Dr. Xavier Sigaud 150, 22290-180, Rio de
Janeiro, Brazil} }

\author{I. S. Oliveira}{
  address={Centro Brasileiro de Pesquisas
F\'{\i}sicas, Rua Dr. Xavier Sigaud 150, 22290-180, Rio de
Janeiro, Brazil} }

\begin{abstract}
During the past few years, nonextensive statistics has been
successfully applied to explain many different kinds of systems.
Through these studies some interpretations of the entropic
parameter $q$, which has major role in this statistics, in terms
of physical quantities have been obtained. The aim of the present
work is to yield an overview of the applications of nonextensive
statistics to complex problems such as inhomogeneous magnetic
systems.
\end{abstract}

\maketitle


\section{Introduction}
The Nonextensive statistics attempts to
handle some anomalies that appear in physical problems which are
not completely described by the usual Boltzmann-Gibbs statistics.
Its applications run through different systems;  magnetic systems
are just one of them. In this work we show that Nonextensive
statistics, due to the agreement between theoretical models and
experiments, is a interesting alternative to first principle
models to the study of complex magnetic systems.

\section{Nonextensive statistics of magnetic systems}
The Tsallis generalized statistics,
or Nonextensive statistics, is based on the definition of the
generalized entropy $S_{q}=
k\left(1-Tr\{\hat{\rho}^{q}\}\right)/(q-1)$, where $k$ is a
positive constant, $q \in \Re$ is the entropic parameter and
$\hat{\rho}$ is the density operator. This entropy functional
recovers the Boltzmann-Gibbs one on the limit $q=1$ ($q\rightarrow
1$). From the maximization of this entropy functional, under a
correct $q$-normalized constraint \cite{RenioPhysA}, is
straightforward to express the density operator as $\hat{\rho} =
\left[1-(1-q)\beta\hat{\mathcal{H}}\right]^{1/(1-q)}/Z_{q}$, in
which $Z_{q}$ is the nonextensive partition function,
$\beta=1/k\,T$ and $\hat{\mathcal{H}}$ is the systems Hamiltonian.
Through the same reasoning, the magnetization of a system in the
nonextensive approach is given by \cite{EPL}
\begin{equation}\label{eq.02}
\mathcal{M}_{q} =
\frac{Tr\left\{\hat{\mu}\,\hat{\rho}^{q}\right\}}{Tr\left\{\hat{\rho}^{q}\right\}}
\end{equation}
where $\hat{\mu}$ is the magnetic moment operator. In Fig.1, it is
illustrated the fit of Eq.(\ref{eq.02}), within the mean field
approximation, to experimental data. On the main graph it is
plotted different values of $T_{c}$, also taken from experimental
data, as a function of $q$, which shows a simple straight line
dependence, $T_{c}^{(q)} = q\,T_{c}^{(1)}$ (For details see Ref.
\cite{EPL}).

\begin{figure}
  \includegraphics[height=.2\textheight]{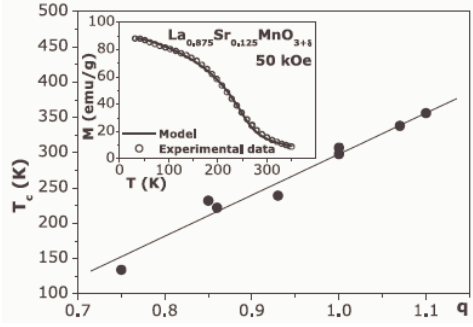}
  \caption{$T_{c}$ versus $q$ shows a linear dependence.}
\end{figure}

Furthermore, one can define the $q$-generalized magnetic moment
thermal average as $\langle\hat{\mu_{z}}\rangle_{q} =
g\,\mu_{B}\,\langle\hat{S}_{z}\rangle_{q} =
q\,\mu_{B}\,S\,\mathcal{B}_{S}^{(q)}$, in which
$\langle\hat{S}_{z}\rangle_{q}$ is the $q$-generalized thermal
average spin operator and $\mathcal{B}_{S}^{(q)}$ is the
\textit{generalized Brillouin function} given by
\begin{equation}\label{eq.03}
\mathcal{B}_{S}^{(q)} =
\frac{1}{S}\langle\hat{S}_{z}\rangle_{q}=\frac{1}{S}
\frac{\sum_{m_{S}=-S}^{+S}m_{S}\left[1+(1-q)\frac{m_{S}}{S}x\right]^{q/(1-q)}}
{\sum_{m_{S}=-S}^{+S}\left[1+(1-q)\frac{m_{S}}{S}x\right]^{q/(1-q)}}
\end{equation}
where $x = g\,\mu_{B}\,\beta\,S\,B$. Using a general expression to
define the magnetic susceptibility such as $\chi_{q} =
\lim_{B\rightarrow 0}\partial_{B}
\langle\hat{\mu_{z}}\rangle_{q}$, one can see in Fig.2 that the
inverse of the susceptibility predicts correctly the behavior of
the paramagnetic systems (For details see Ref. \cite{PRB02}).

\begin{figure}
  \includegraphics[height=.2\textheight]{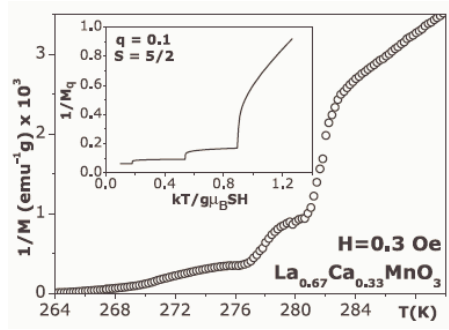}
  \caption{Inverse of the susceptibility versus temperature for a paramagnetic system. The inset represents the
same quantity calculated for a nonextensive magnetic model where
$q = 0.1$ and spin value $S = 5/2$}
\end{figure}

On the other hand, considering a classical spin $\vec{\mu}$ under
a homogeneous magnetic field $\vec{H}$, whose Hamiltonian is given
by $\mathcal{H} = -\mu\,H\,\cos\theta$, it is then possible to
determine the \textit{generalized Langevin function} from
Eq.(\ref{eq.02}), which yields:
\begin{equation}\label{eq.04}
\mathcal{L}_{q} =
\frac{\mu}{(2-q)}\left[\coth_{q}\left(\frac{\mu\,H}{k\,T}\right)-\frac{k\,T}{\mu\,H}\right]
\end{equation}
in which $\coth_{q}$ is the $q$-hyperbolic cotangent. This
expression can be obtained also taking the limit of large spin
values ($S\rightarrow\infty$) on the generalized Brillouin
function. As can be seen on Fig.3, this magnetization, also within
the mean field approximation, predicts a first-order type phase
transition for $q < 0.5$ and a second-order type for $q
> 0.5$ which which occurs in certain kinds of magnetic systems (For
details see Ref. \cite{PRB03}).

\begin{figure}
  \includegraphics[height=.2\textheight]{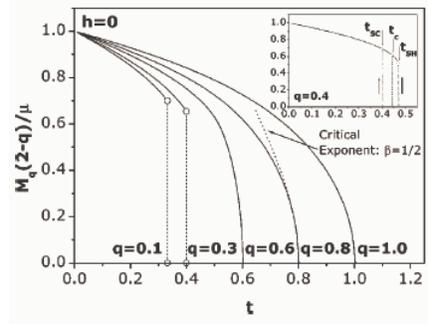}
  \caption{Magnetization versus temperature for several values of $q$ and $h=0$. For $q<0.5$ the
transition is of first-order type, whereas for $q>0.5$ the
transition is of second-order type.}
\end{figure}

\section{Interpretation of the entropic parameter}
Now, with these magnetic quantities defined in a nonextensive
approach, one may ask the question: If is possible to connect a
homogenous nonextensive system to a inhomogeneous one described by
the Boltzmann-Gibbs statistics could the entropic parameter be
interpreted through some magnetic quantities?

So, to answer this question, lets us consider an inhomogeneous
system (described by the Boltzmann-Gibbs statistics) composed by
magnetic clusters distributed in size, therefore in their net
magnetic moment, and the $f(\mu)$ representing this distribution.
The average magnetization will be given by
\begin{equation}\label{eq.05}
\langle\mathcal{M}\rangle =
\int_{0}^{\infty}\mathcal{M}\,f(\mu)\,d\mu
\end{equation}
From Eqs.(\ref{eq.04}) and (\ref{eq.05}), the average and the
nonextensive susceptibilities can be derived as
$\langle\chi\rangle = \langle\mu^{2}\rangle/3kT$ and $\chi_{q} =
q\mu^{2}/3kT$ respectively. Calculating saturation values of the
magnetization one finds $\langle\mathcal{M}\rangle_{sat} =
\langle\mu\rangle$ for the average and $\mathcal{M}_{q,sat} =
\mu/(2-q)$ for the nonextensive. Thus, equating those limits
($\langle\chi\rangle = \chi_{q}$ and
$\langle\mathcal{M}\rangle_{sat} = \mathcal{M}_{q,sat}$), a
microscopic analytical expression to the $q$ parameter is found
\begin{equation}\label{eq.06}
q\,(2-q)^{2} = \frac{\langle\mu^{2}\rangle}{\langle\mu\rangle^{2}}
\end{equation}
where $\langle\mu\rangle$ and $\langle\mu^{2}\rangle$ are the
first and second moments of the distribution $f(\mu)$. This result
is valid for any distribution $f(\mu)$ and gives to the entropic
parameter an interpretation, which involves magnetic quantities,
that can experimentally determined (For details see
Refs.\cite{PRB06, EPJB}).

However, it is possible to try a more fundamental magnetic
quantity considering a ferromagnetic $N$ spins system. In this
case, let consider that each spin interacts with $z$ neighbors in
an inhomogeneous medium, in the Boltzmann-Gibbs framework
\cite{JSTAT}. As the interactions are inhomogeneous, one must
consider a distribution $f(J)$ of the system exchange integrals on
the system. Thus, the average magnetization variation per unit of
volume is given by
\begin{equation}\label{eq.07}
\langle\Delta\,m\rangle =
\frac{\zeta(3/2)g\mu_{B}}{8\pi^{3/2}}\left(\frac{k_{B}T}{a^{2}\,z\,S}\right)^{3/2}\int_{0}^{\infty}dJ\frac{f(J)}{J^{3/2}}
=\frac{\zeta(3/2)g\mu_{B}}{8\pi^{3/2}}\left(\frac{k_{B}T}{a^{2}\,z\,S}\right)^{3/2}\langle
J^{-3/2}\rangle
\end{equation}
Now, calculating the same quantity but on a homogeneous
nonextensive system one can find \cite{JSTAT}
\begin{equation}\label{eq.08}
\langle\Delta\,m\rangle_{q} =
\frac{g\mu_{B}}{2\,\pi}\int_{0}^{\infty}dk\,k^{2}\,\frac{Tr\{n_{k}\rho^{q}\}}{Tr\{\rho^{q}\}}=
\frac{g\mu_{B}}{4\,\pi}
\left(\frac{k_{B}\,T}{a^{2}\,S\,\mathcal{J}}\right)^{3/2}F(q)
\end{equation}
where $F(q)$ is a dimensionless integral that depends only on the
value of the entropic parameter $q$. From the mean field
approximation it is then possible to find the critical temperature
$T_{c}^{(1)} = z\,S(S+1)\langle J\rangle/3k_{B}$ for the
inhomogeneous Boltzmann-Gibbs and $T_{c}^{(q)} =
z\,S(S+1)\,q\,\mathcal{J}/3k_{B}$ for the homogeneous
nonextensive. Using the relation showed above between these two
critical temperatures, one can find that $\mathcal{J} = \langle
J\rangle$. Thus, comparing Eqs.(\ref{eq.07}) and (\ref{eq.08}),
another expression is obtained
\begin{equation}\label{eq.09}
F(q) =
\frac{\sqrt{\pi}}{2}\zeta\left(\frac{3}{2}\right)\frac{\langle
J^{-3/2}\rangle}{\langle J\rangle^{-3/2}}
\end{equation}
revealing a relation between the $q$ parameter and moments of the
exchange interaction distribution $f(J)$ (For details see
Ref.\cite{JSTAT}).

\section{Conclusions}
Summarizing, we showed that the nonextensive statistics is an
useful alternative for the description of complex behavior of some
kinds of magnetic systems. It was also showed that the entropic
parameter $q$ can be interpreted through magnetic quantities, and
can be seen as a  measurement of the inhomogeneity of a magnetic
system.




\begin{theacknowledgments}
The authors would like to thanks the brazilian funding agencies
CNPq and CAPES. DOSP would like to thanks the brazilian funding
agency CAPES for the financial support at Universidade de Aveiro
at Portugal.
\end{theacknowledgments}





\IfFileExists{\jobname.bbl}{}
 {\typeout{}
  \typeout{******************************************}
  \typeout{** Please run "bibtex \jobname" to optain}
  \typeout{** the bibliography and then re-run LaTeX}
  \typeout{** twice to fix the references!}
  \typeout{******************************************}
  \typeout{}
 }



\end{document}